# Entangled photon triplets using lithium niobate nanophotonics

Nathan A. Harper[1]†, Ayantika Sengupta[2]†, Emily Y. Hwang[2], Scott K. Cushing[1]*

[1]Department of Chemistry and Chemical Engineering, California Institute of Technology; Pasadena, California 91125, USA.

[2]Department of Applied Physics and Materials Science, California Institute of Technology; Pasadena, California 91125, USA.

*Corresponding author. Email: scushing@caltech.edu

†These authors contributed equally to this work.

**Abstract:** Multiphoton states are needed for quantum communication and computation. Multiphoton states are significantly more difficult to generate than one- and two-photon states because two individual down-conversion processes must be cascaded. Only efficiencies of <100 Hz/mW have been reported to date. We integrate two down-converters on the same thin-film lithium niobate waveguide, significantly enhancing the cascaded process efficiency to 237 ± 36 kHz/mW. The measured $4.4 \times 10^{-5}$ probability of the second down-converter, which sets the limit on detectable triplet rates, exceeds those of previous triplet sources by an order of magnitude and demonstrates a path towards MHz rates of triplets for quantum applications.

States with more than two entangled photons are needed in quantum communication and computation, amongst other applications (*1-4*). Emitters such as atoms and quantum dots can generate single photons on demand (*6-7*). However, efficient photon collection is challenging, and qubit dephasing limits state fidelity during each emission cycle. Entanglement swapping is the method of choice for generating multiphoton entanglement because spontaneous parametric down-conversion (SPDC) has a high state fidelity (*8-12*). In entanglement swapping, independent SPDC sources are combined interferometrically to erase the knowledge of which source produced each photon. SPDC, however, is not deterministic. Multiphoton generation in entanglement swapping can only be determined after the fact by counting the photons generated. This measurement might prevent further use of the state.

Cascaded SPDC (CSPDC) directly produces multiphoton states from one pump photon (*14-21*). A pump photon is first down-converted into a pair of daughter photons. One of the daughter photons is then down-converted using a second SPDC interaction to produce a pair of granddaughter photons. All three photons exist in a time-energy entangled state because both SPDC processes conserve energy (*15*). The produced entangled photon pair can be heralded by detecting one of the granddaughter photons. The remaining two photons can now be used. Three independent SPDC sources (with a four-fold detection event) would be required to replicate the same result in entanglement swapping (*22-24*).

However, CSPDC has had limited practical applications because the detected triplet rates are <1 Hz. The limiting parameter is the efficiency of the second down-converter. The flux of granddaughter photons is orders of magnitude lower than the flux of daughter photons. The pump power cannot be arbitrarily increased because the detector for daughter photons saturates well before the granddaughter detectors. To date, the second down-conversion process relies on transferring pairs from the first down-converter to a long (>3 cm) diffused or micromachined periodically poled lithium niobate (PPLN) waveguide. Significant losses (>50%) occur in the transfer process. The down-converters are also limited in efficiency by their large modal areas. They approach practical limits in the interaction lengths they can achieve, set by the dimensions of the precursor material.

Thin-film lithium niobate (TFLN) waveguides with sub-micron modal areas have significantly improved the efficiency of nonlinear interactions despite short interaction lengths (<1 cm) (*25-27*). We use TFLN to improve CSDPC rates to 237 ± 36 kHz/mW, orders of magnitude higher than previous reports. Integrating two down-converters onto the same waveguide allows transfer of generated pairs to the second down-converter with low loss (Fig. 1). In addition to tight modal overlaps to scale the brightness of each source, the efficiency of the second down-converter is further improved by tuning the waveguide dispersion through the device geometry (*28, 29*). Working on a low-loss integrated platform also allows for significantly longer interaction lengths, either through high quality factor cavities or by folding the waveguide to achieve path lengths much longer than the physical dimensions of the substrate (*30, 32*). This provides a new route to scaling down-converter efficiency that is not possible in diffused/micromachined large-area waveguides.

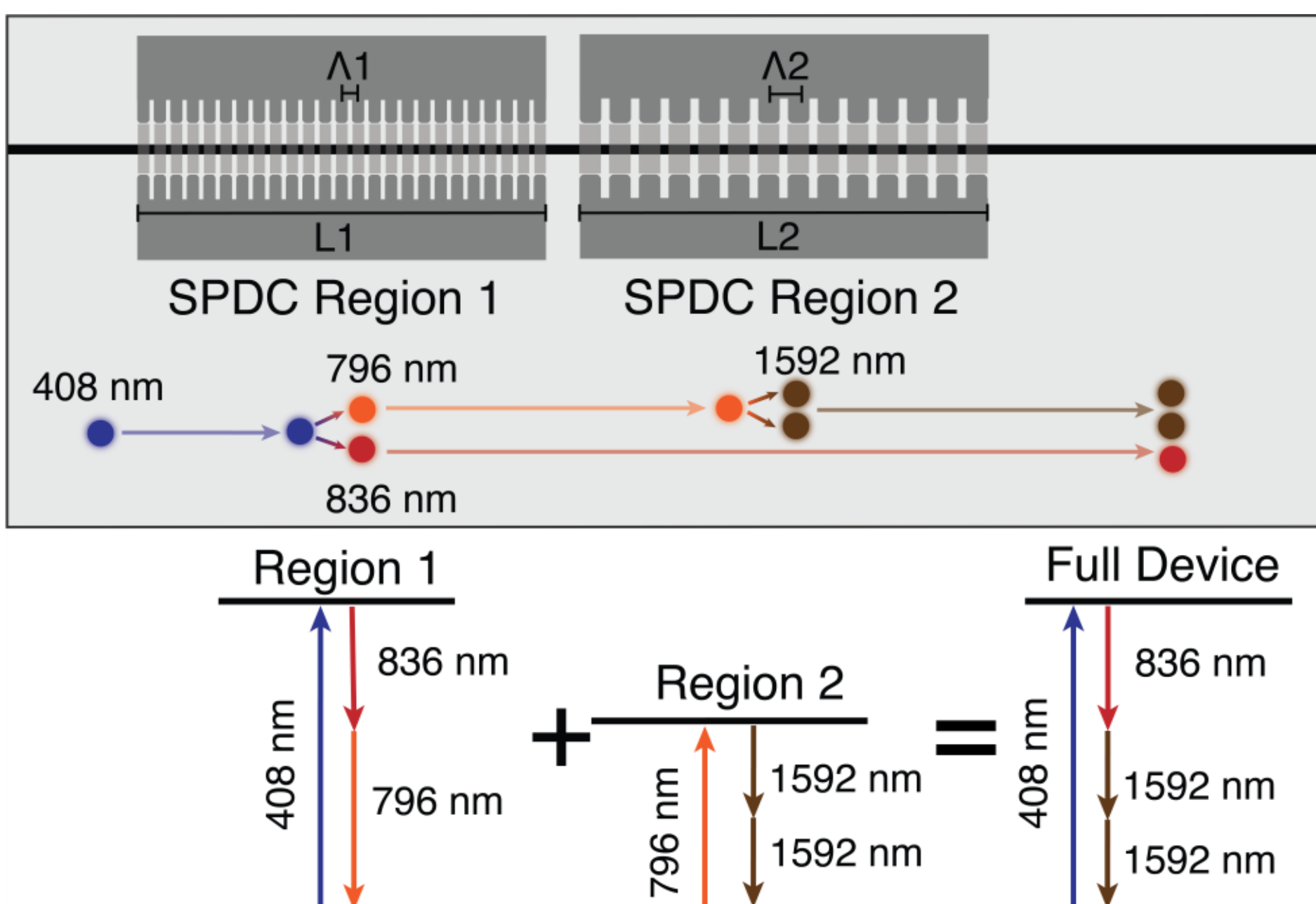

**Fig. 1. Conceptual overview of the CSPDC device and process.**
(**Top**) Schematic of the TFLN-based waveguide containing two periodically poled down-conversion regions. In the first region, a pump produces a pair of daughter photons through SPDC; one daughter photon acts as a pump for the second down-converter and produces a pair of granddaughter photons. All photons are confined in one waveguide. L, down-converter length; Λ, poling period. (**Bottom**) CSPDC process diagrams showing the center wavelengths of all photons involved, as set by energy conservation and phase matching.

The device is pumped with a continuous-wave violet laser, which produces a pair of daughter photons centered at 796 nm and 836 nm. The daughter photons are equally spaced from a degenerate wavelength of 816 nm. The daughter photons travel through a second down-converter, which is phase-matched for broadband SPDC in the telecom range. Three photons are produced in the waveguide, consisting of the 836 nm daughter photon and two granddaughter photons spanning 1300-2100 nm (centered at a degenerate wavelength of 1592 nm). The off-chip detection of one granddaughter photon certifies that the two interactions were successful, heralding the presence of the remaining entangled daughter (836 nm) and granddaughter photon.

The detected triplet rate, with known on-chip input pump power ($R_{pump}$), individual efficiencies of the two down-converters ($E_1$ and $E_2$, respectively), and overall detection efficiencies can be written as

$$R_{triplet} = R_{pump} E_1 E_2 \eta_{836} \eta_{1592}^2 + Acc_{triplet} \quad (1)$$

where $\eta_{836}$ and $\eta_{1592}$ denote the total transmission of the daughter and granddaughter photons, respectively, from their point of generation to detection. On-chip propagation loss, collection efficiency, and detector efficiency are included. We assume that all 796 nm pump photons generated at a rate $R_{pump} E_1$ are transmitted to the second down-converter, which is valid for a short separation on a low-loss platform, such as TFLN. $Acc_{triplet}$ is the rate of accidental triplet detections within a coincidence window $\tau$. This accidental term contains contributions from (i) events where two photons from a genuine triplet are detected in coincidence and a third, uncorrelated photon is detected on the remaining detector, and (ii) events where all three detected photons are mutually uncorrelated. In a low-gain regime ($R_{pump} E_1 \tau \ll 1$), the dominant contribution comes from a true two-fold coincidence detection between granddaughter photons with an uncorrelated photon (dark count or background event) detected at the daughter detector. This is because the daughter arm has a much higher singles rate than the granddaughter arms, so the probability that a noise count falls within the triplet coincidence window is significantly higher. The accidental term can be approximated as

$$Acc_{triplet} \approx (R_{pump} E_1 E_2 \eta_{1592}^2)(R_{836} \tau) \quad (2)$$

Here, $R_{836}$ is the rate of singles at the daughter detector. The first bracket gives the rate of true two-fold coincidences between the granddaughter photons. In contrast, the second bracket is the probability that an additional uncorrelated photon is detected at the daughter detector within the same window. It is important to note that the above assumption breaks down if unwanted phase matching of higher-order pump modes in the first down-converter generates idler photons in the telecom band. The photons then appear as noise in the granddaughter channels, and their rate scales with pump power. The accidental contribution from (ii) can become comparable to the term in Eq. 2 as the pump power is increased. Assuming detector dark counts to be negligible as compared to true daughter counts from SPDC, $R_{836}$ increases linearly with $E_1$. Thus, the product of the triplet coincidence-to-accidental (CAR) ratio (from Eqs. 1 and 2) and the on-chip triplet generation rate ($R_{pump} E_1 E_2$) approximately scales linearly with the efficiency of the second down-converter, making it a critical performance metric of a CSPDC device.

The cascaded down-conversion device was fabricated using standard thin-film lithium niobate processes (Fig. 2A) (*25*). Two down-converters of equal length (3.5 mm) were produced with different quasi-phase matching periods tailored for the first and second down-conversion processes, as shown in the two-photon microscopy image (Fig. 2B). The first down-converter uses a period of 2.1 μm, which satisfies first-order Type-0 quasi-phase matching between the fundamental TE modes of the 408 nm pump, 796 nm signal, and 836 nm idler. As this device uses X-cut TFLN (crystal axis polarized in the plane of the thin film), the strongest nonlinear tensor element of lithium niobate ($d_{33}$) is used. There is excellent agreement in the shape of the spectrum between experiment and theory (Fig. 2C), suggesting a uniform duty cycle and negligible variations in the waveguide parameters along the length of the down-converter. A brightness of 3.5 GHz/mW/nm at 796 nm is experimentally observed, making this one of the brightest single-pass down-converters demonstrated to date. The measured brightness is approximately two times lower than the theoretical calculation, potentially due to an overestimation of the pump coupling efficiency into the waveguide or an underestimation of the propagation losses at the pump wavelength. While a highly efficient first down-converter is not strictly necessary to achieve high triplet rates (assuming sufficient laser power is available), the ability to operate at low pump powers is undoubtedly beneficial for reducing autofluorescence, photorefraction, and thermal instabilities, all of which are potential issues at near-UV wavelengths in nonlinear crystals.

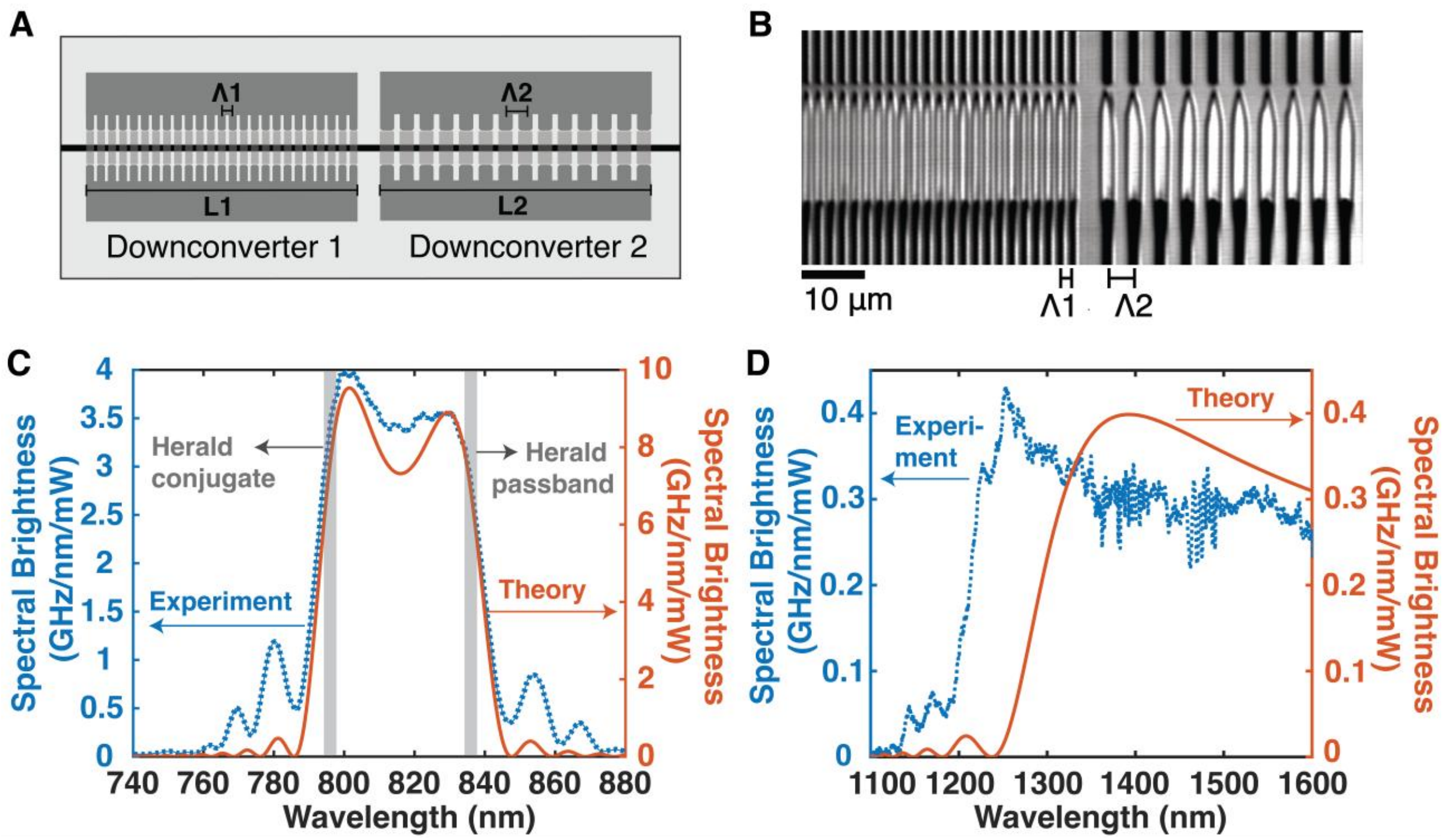

**Fig. 2. Cascaded down-conversion device and spectral characterization.**
(**A**) Schematic of the CSPDC device. (**B**) A second harmonic microscope image of the cascaded down-conversion regions with different poling periods to satisfy type-0 phase matching for the two different combinations of wavelengths. Measured (left axis) and theoretical (right axis) spectral brightness of the (**C**) first and (**D**) second down-converter. The grey strips in (**C**) highlight the wavelengths for pumping the second down-converter and the corresponding heralding wavelengths for CSPDC.

The second down-converter is designed for first-order Type-0 phase matching at the 796 nm pump and 1300-2000 nm signal and idler photons to maximize the brightness of the interaction. Like many previous works (*29, 30*), the bandwidth of the down-conversion interaction is increased through dispersion engineering. By tailoring the waveguide top width to minimize the group velocity dispersion of the fundamental TE mode (calculated to be -8 $fs^2$/mm at 1592 nm), a very broad down-conversion spectrum is experimentally achieved (Fig. 2D). While wavelengths beyond 1650 nm are outside the range of the spectrometer used here, a 3-dB bandwidth of 116 THz (1218-2299 nm) is inferred from the signal photons. The brightness of the second down-converter is lower than the brightness of the first down-converter due to the longer wavelengths of the fields, as well as the lower $d_{33}$ strength at these wavelengths, which is also captured in the theoretical spectrum. When integrating the brightness from 1300 nm (the cutoff wavelength of the long pass filters used in the subsequent experiments) to 1592 nm (the degenerate wavelength) and multiplying by a factor of two (accounting for symmetry of the signal and idler), an efficiency of 175 GHz/mW is inferred. This corresponds to a down-conversion probability of $4.4 \times 10^{-5}$ for 796 nm photons in the waveguide, which is also among the most efficient down-converters demonstrated at this wavelength range. Both down-converters co-integrated onto the same waveguide are therefore among the best in their class. Full details on the characterization of the brightness, spectra, and theory for both down-converters are available in the Supplementary Materials.

With both down-converters individually characterized, their cascaded performance to generate entangled photon triplets was measured using the setup shown in Fig. 3A. The device is pumped with a single frequency CW laser at 408 nm, and the generated daughter and granddaughter photons are collected off-chip using a high numerical aperture aspheric lens (NA=0.5). The daughter and granddaughter photons are then spectrally separated with a dichroic mirror and sent to Si and InGaAs single photon avalanche detectors (SPADs), respectively. Coincidences between the detection of the daughter photon and one granddaughter photon are used to directly infer the triplet production rate using a time-correlated single photon counting board. The only filtering on the granddaughter photon detector is a set of two long-pass filters with a 1300 nm cutoff wavelength, and the longest wavelengths that are effectively detected are limited by the spectral response of the InGaAs detector (approximately 1600 nm). More stringent filtering is required on the daughter arm because the SPDC bandwidth of the first down-converter (46 nm FWHM) is much broader than the acceptance bandwidth of the second down-converter (3.7 nm FHWM). A 3 nm bandpass filter is therefore present in the daughter arm, allowing only the photons most likely to correspond with a triplet to be detected. A wavelength of 836 nm for these idler photons is conjugate to the 796 nm pump wavelength of the second down-converter. The filter was tuned to this wavelength by rotating the filter on a rotation stage. Two additional long pass filters (cutoff wavelengths of 700 nm and 808 nm) were used to suppress any residual 408 nm light.Figure. 3B shows the coincidence histogram recorded over a 16-hour acquisition time using this setup, showing a coincidence peak due to the correlated emission of daughter and granddaughter photons from the device. A coincidence rate of $0.20 \pm 0.03$ Hz is estimated with 188 nW of on-chip pump power after subtracting the accidental baseline. The probabilities of detecting a generated herald daughter photon and a granddaughter photon are estimated as 13.8% and 3.3%, respectively (Supplementary Materials). The resulting triplet generation efficiency and rate are therefore $237 \pm 36$ kHz/mW and $45 \pm 7$ triplets/s, respectively. This rate agrees well with the rate predicted from the separate characterization of the down-converters in Fig. 2 using Eq. 3:

$$E_{CSPDC} = B_{796} \cdot BW_{836} \cdot E_2 \cdot \eta \tag{3}$$

Here $B_{796}$ is the brightness of the first down-converter (at 796 nm), $BW_{836}$ is the effective bandwidth of the bandpass filter on the daughter arm, $E_2$ is the per-photon efficiency of the second down-converter, and $\eta$ is a constant which quantifies the reduction in efficiency of the second down-converter when pumped with a spectrally broadband source instead of a single frequency laser. The effective filter bandwidth is defined as the width of a flat-topped ideal filter that gives the same integrated transmission as the actual filter used here, conjugated to 796 nm, and is estimated to be 4.1 nm (Supplementary Materials). $\eta$ is estimated from theory to be 0.83 and depends on the width of the bandpass filter and group velocity mismatch of the second down-converter. Using a brightness $B_1$ of 3.5 GHz/mW/nm and efficiency $E_2$ of $4.4 \times 10^{-5}$, a predicted triplet efficiency of 524 kHz/mW is inferred. This disagrees with the experimental result by approximately a factor of two. The discrepancy likely arises from uncertainty in the estimated on-chip power of the 796 nm daughter photons due to propagation losses between the two nonlinear sections, as well as uncertainties in the input and output coupling efficiencies. Moreover, the ultra-broadband granddaughter photons are detected at SPADs with a non-uniform spectral response over their sensitivity range, which reduces the spectrally averaged detection efficiency as compared to the individual brightness measurement of down-converter 2, where the telecom photons were detected with a 30 nm bandpass filter centered at 1550 nm (Supplementary Materials).

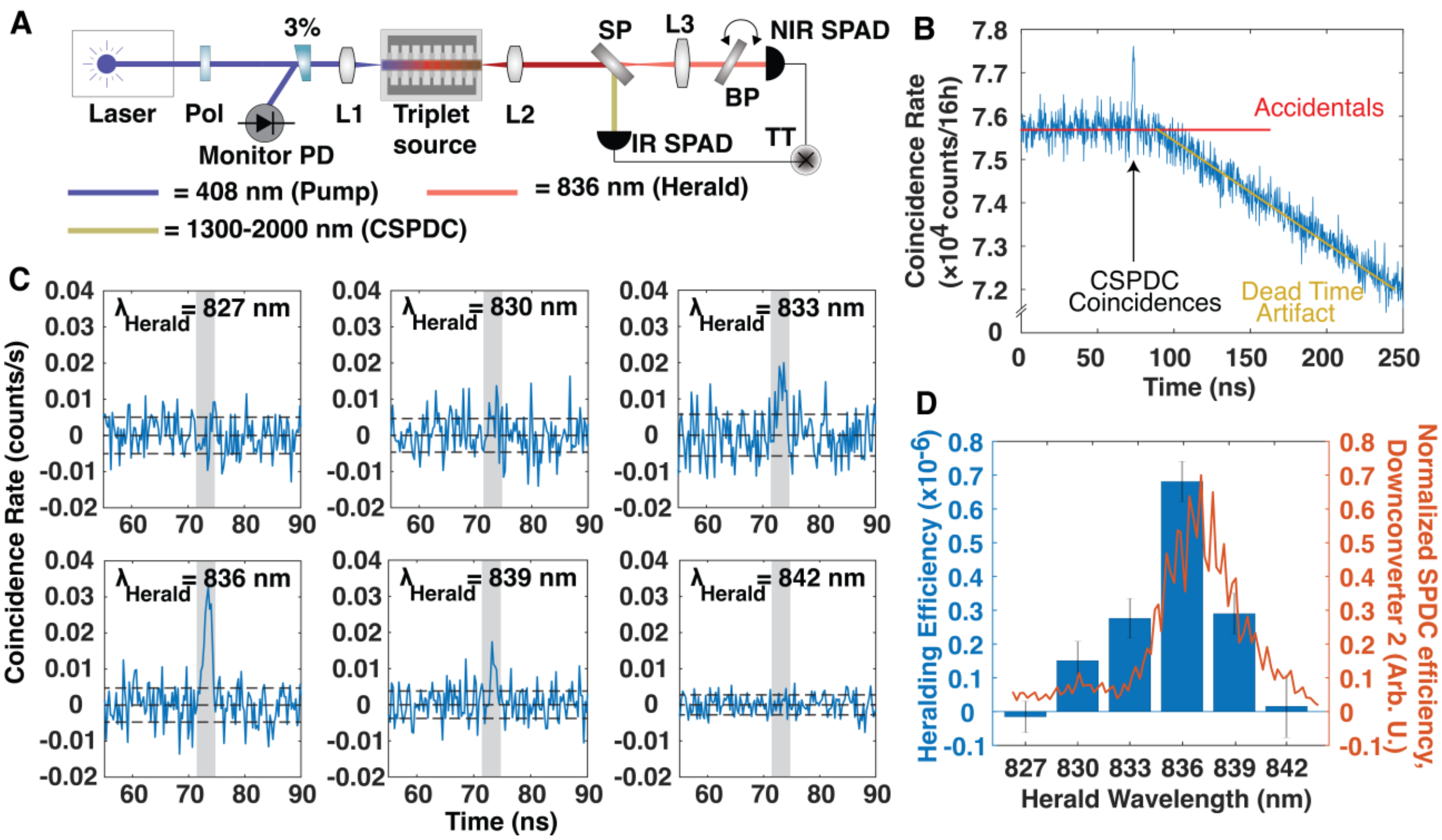

**Fig. 3. Combined performance of the CSPDC device.**
(**A**) Simplified setup for measuring CSPDC generation efficiency. For clarity, a pump half-wave plate and a 500 nm short-pass filter are not shown, nor are mirrors or the long-pass filters in front of the NIR and IR SPAD fiber couplers. The collection aspheric lens (L2) collimates the telecom-band output while the 836 nm photons are converging in free space and are collimated using an achromatic lens (L3). Abbreviations: Pol., polarizer; PD, photodiode; SP, dichroic short-pass ($\lambda$ = 1180 nm); BP, angle-tuned bandpass ($\lambda$ = 836 nm); TT, time tagger. (**B**) Coincidence histogram recorded over 16 hours without accidental subtraction. The histogrammer resolution is 4 ps, and post-processed into 256 ps bins. A decrease in coincidences after $\approx$ 90 ns is observed due to histogrammer dead time. (**C**) Accidental-subtracted coincidence histograms at six heralding wavelengths. Gray regions show the time bins used for integrating coincidences, and dotted lines show the 1$\sigma$ uncertainties of the accidental floor. (**D**) Heralding probabilities of CSPDC events (blue bars, left axis) and the relative efficiency of Down-converter 2 at 1550 nm for different pump wavelengths, converted to conjugate wavelengths by energy conservation (red line, right axis).

To confirm that the origin of the coincidence peak in Fig. 3B is due to CSPDC, a control experiment was performed where the angle of the bandpass filter was detuned from 836 nm. Fig. 3C shows coincidence histograms recorded at different central wavelengths of the filter. As the herald wavelength shifts, its partner photon moves outside the acceptance bandwidth of the second down-converter, leading to fewer generated granddaughter photons, and a lower coincidence rate. A bar graph summarizing the estimated probability of detecting a coincidence when a photon is present in the herald arm is shown in Fig. 3D. The heralding probability agrees well with the pump acceptance spectrum of the second down-converter.

The primary challenge in detecting triplets in our setup is the high noise floor in the coincidence histogram, which necessitated long integration times despite the intrinsically high triplet generation rate. This background arises from accidental coincidences between uncorrelated herald photons and noise photons within the sensitivity range of the IR SPADs. While the expected count rate from CSPDC photons is only ~3.1 cps, the IR SPADs registered ~17 kcps, indicating the presence of photons not originating from CSPDC. To identify its origin, we measured the SPDC spectrum of the device over the wavelength range conjugate to the SPAD sensitivity band (535–590 nm). Multiple peaks were observed in this region, confirming the presence of corresponding idler photons between 1300–1700 nm generated by higher-order quasi-TE pump modes satisfying first-order phase matching in the first down-conversion region (Supplementary Materials). These modes generate highly non-degenerate SPDC, with quasi-TE idler modes falling within the SPAD detection range. Future designs can mitigate this by incorporating mode-filters using tapered waveguides to suppress higher-order pump modes (*33*). Alternatively, a short-pass filter at 1300 nm before the second down-converter could be used to prevent these unwanted IR photons from interfering with the true triplet signal. A simpler approach is to place a free-space notch interference filter after the device to block the spurious idler peaks, with the trade-off of some CSPDC loss. Finally, modifying the waveguide geometry could spectrally shift the parasitic SPDC peaks away from the IR SPAD detection band.

CSPDC offers a promising route for the generation of multiphoton entanglement and heralded generation of time-energy and time-bin entangled Bell states, when combined with time-bin encoding. Compared to six-photon schemes that require emission of three SPDC pairs and multiphoton interference using linear optics to herald a single pair, our integrated cascaded platform naturally yields time–energy entangled photon triplets, without stringent requirements on interferometric stability or photon indistinguishability between independent SPDC sources. The two key figures of merit of a CSPDC device are (i) the transmission of the daughter pump photon produced in the first region to the second and (ii) the down-conversion efficiency of the second stage. Previous CSPDC demonstrations using separate nonlinear platforms have been limited by coupling losses at the interface between the two stages. In our architecture, the daughter photons are confined on-chip, therefore, the transfer probability is limited only by propagation loss between the two regions, which on TFLN can be as low as 0.06 dB/cm at visible wavelengths (*32*). The efficiency of the second SPDC, which limits the detectable triplet rate, can be boosted by using very long diffused waveguides; however, increasing the length of the nonlinear region also narrows the acceptance bandwidth, reducing the effective usable output of the first down-converter. Our dispersion-engineered second down-converter, only 3.5 mm long, achieves an efficiency comparable to that of 30-mm diffused PPLN crystals used in prior CSPDC work.

Table 1 summarizes the key metrics of comparable experiments and our work. Our device achieves a triplet generation efficiency three to four orders of magnitude higher than previous demonstrations. This substantial improvement stems from a simple yet powerful architectural choice: all photons remain confined to a single waveguide throughout the cascaded process. By leveraging TFLN's low propagation loss and tight modal confinement, our dispersion-engineered waveguide enables highly efficient second down-conversion while maintaining a compact device footprint. Increasing the detection rate requires further increasing the efficiency of the second down-converter. Adapted poling to combat index variations of the 3.5 mm devices could potentially reach the theoretical performance of such devices, but would represent a modest 35% improvement (*34*). Longer down-converters could be used instead, which will require adapted poling since index variations become more severe over longer distances. A potential solution is a meandering waveguide, where the efficiency of the second down-converter scales with the number of meanders, but is ultimately limited by the 796 nm daughter photon's propagation losses. Alternatively, an approach that uses a resonator to repeatedly circulate the 796 nm daughter photon through the same Down-converter 2 could be very promising (*35*). Finally, the IR SPADs used here have only ~20% detection efficiency at 1550 nm. Replacing them with SNSPDs, which can achieve >90% efficiency (36), would boost the detectable triplet rate by more than an order of magnitude. This improvement would increase the coincidence SNR and reduce the required integration time from hours to practical timescales.

| Ref. | Source 1 | Source 2 | $\eta_{1,2}$ | $\eta_{DC2}$ | $\eta_{1,2} \times \eta_{DC2}$ | Triplet rate (Hz) | Efficiency (Hz/mW) |
|---|---|---|---|---|---|---|---|
| (*14*) | PPKTP bulk crystal | PPLN waveguide | 0.3 | $1 \times 10^{-5}$ | $2.6 \times 10^{-6}$ | 4.3* | 1.8* |
| (*15*) | PPKTP bulk crystal | PPLN waveguide | 0.3 | $1 \times 10^{-5}$ | $2.6 \times 10^{-6}$ | 0.8* | 0.06* |
| (*16*) | PPKTP bulk crystal | PPLN waveguide | 0.5 | $7 \times 10^{-6}$ | $3.5 \times 10^{-6}$ | 8.4* | 0.3* |
| (*17*) | PPKTP bulk crystal | PPLN waveguide | 0.3 | $1 \times 10^{-6}$ | $3 \times 10^{-7}$ | 0.3* | 0.03* |
| (*18*) | PPKTP bulk crystal | PPLN waveguide | 0.18* | $1 \times 10^{-6}$ | $2 \times 10^{-7}$ | 2.6* | 0.6* |
| (*19*) | PPLN waveguide | PPLN waveguide | ≈ 1 | $3 \times 10^{-7}$ | $3 \times 10^{-7}$ | 0.58* | 57.6* |
| (*20*) | Hot $^{85}$Rb atomic vapor (SRS) | PPLN waveguide | – | $1 \times 10^{-6}$ | – | 320.8 | 2.5 |
| **This work** | **TFLN waveguide** | **TFLN waveguide** | **≈ 1** | $\mathbf{4 \times 10^{-5}}$ | $\mathbf{4 \times 10^{-5}}$ | **45** | $\mathbf{1.3 \times 10^{5}}$ |

**Table 1. Summary of relevant metrics of comparable CSPDC experiments.**
Values that were not directly given and were indirectly calculated are indicated with *. Values that could not be reliably inferred are indicated with –. In this table, 'PPLN waveguide' denotes a large-area PPLN waveguide. $\eta_{1,2}$: Transmission efficiency from Down-converter 1 to Down-converter 2, $\eta_{DC2}$: Down-converter 2 per-photon efficiency, SRS: Spontaneous Raman scattering.

**Acknowledgments:** The authors gratefully acknowledge the critical support and infrastructure provided for this work by The Kavli Nanoscience Institute (KNI) and the Beckman Biological Imaging Facility at Caltech. The authors thank Ryoto Sekine and Alireza Marandi for providing the periodic poling apparatus and for helpful discussions.

**Funding:** This work was funded by the Defense Advanced Research Projects Agency Young Faculty Award (YFA) program (D24AP00312).

**Author contributions:** NAH and EYH conceived of the idea. EYH designed and fabricated the device. NAH and AS designed and carried out the experiments. NAH and AS performed the theoretical and data analysis. NAH and AS wrote the manuscript with input from all other authors; and SKC supervised the project.

**Competing interests:** Authors declare that they have no competing interests.

**Data and materials availability:** All data needed to evaluate the conclusions in this paper are available in the main text and Supplementary Materials. Raw data are available from the corresponding author upon reasonable request.

# Entangled photon triplets using lithium niobate nanophotonics: Supplementary Materials

**Cascaded spontaneous parametric down-conversion (CSPDC) device design**

The cascaded device consists of two periodically poled regions on an X-cut 1 cm × 1 cm thin film lithium niobate (TFLN) on insulator chip (NanoLN). The effective modal refractive indices and dispersion properties of the waveguide geometry shown in Fig. S1A are simulated on a finite-difference eigenmode solver software using the bulk Sellmeier coefficients of 5% MgO-doped lithium niobate and $SiO_2$. The fundamental quasi-TE modes are considered for type-0 phase matching at the pump and the spontaneous parametric down-conversion (SPDC) signal and idler wavelengths of the two nonlinear regions to leverage the highest nonlinear coefficient in lithium niobate ($d_{33}$). The waveguide top width is optimized to minimize the group velocity dispersion (GVD) near the degenerate wavelength of 1592 nm for the second SPDC process. The GVD as a function of the waveguide top width and etch depth is shown in Fig. S1B. Poling periods for the two interactions are computed using the simulated modal data for the final waveguide design. The length of both the poled regions is 3.5 mm each. The waveguide fabrication process is described in detail in Ref. (*25*). The post-fabrication device geometry is imaged using atomic force microscopy, indicating a top width of 2.12 μm, a film thickness of 610.9 nm, an etch depth of 520 nm and a sidewall angle of 62°. The facets of the chip are polished for efficient optical coupling, resulting in a total waveguide length of 8 mm. The GVD as a function of wavelength for the measured geometry is shown in Fig. S1C. The near-zero GVD ($-8$ fs$^2$/mm) at 1592 nm indicates broadband SPDC in the second down-converter, consistent with experimental results.

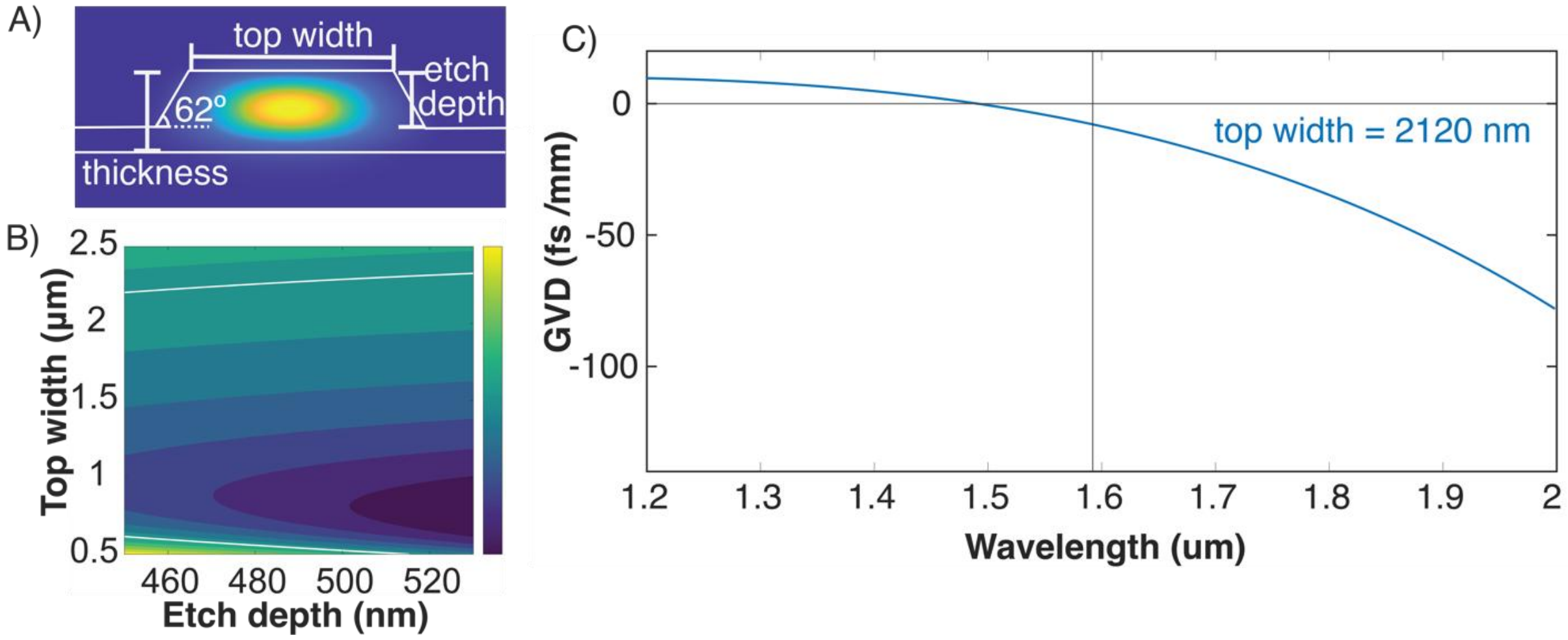

**Fig. S1.** CSPDC device design. (A) Waveguide cross-section and simulated mode profile. (B) GVD as a function of waveguide top width and etch depth. The white lines correspond to GVD = 0. (C) Simulated GVD for our waveguide of 2.12 μm top width, 520 nm etch depth and 610.9 nm thickness. A vertical line at 1592 nm is shown.

**Characterization of the cascaded device and individual down-converters**

All experiments are driven by a widely tunable, single-frequency continuous-wave (CW) Ti:sapphire laser. The laser wavelength is stabilized against long-term drifts using a wavemeter.

Depending on the measurement, the waveguide is pumped either at 408 nm or 796 nm, corresponding to the pump wavelengths of the first and second down-converter, respectively. A variable neutral density filter controls the waveguide input pump power. The 408 nm pump beam is generated by frequency doubling the output of the laser at a BBO crystal and is coupled into a single-mode polarization maintaining fiber (PMF) after filtering the fundamental harmonic using an OD 5 shortpass filter (cutoff λ = 500 nm). The fiber carries the pump beam to another optical table where the chip-coupling setup is located. The collimated fiber output ($1/e^2$ beam diameter ≈ 1 mm) is further filtered to remove any fluorescence generated in the fiber. A photodiode sensor records the amount of free space pump power in the experiments using the reflection from an uncoated wedge. The collimated 408 nm pump passes through a polarizer and a half-wave plate to prepare linear polarization along the optic axis of the thin-film lithium niobate (TFLN) chip. The beam is then focused onto the input facet of the waveguide with a 0.6 NA, 1.5 mm focal length aspheric lens (Thorlabs C140TMD-A), forming an estimated spot size of 775 nm. A similar beam-delivery and fiber-to-free-space-coupling setup is built for using the fundamental harmonic of the laser at 796 nm to specifically pump the second down-converter, followed by spectral and polarization conditioning.

The chip studied here is maintained at a temperature of 25°C by a crystal oven. Over 150 waveguides are present with varying top widths and poling periods. The specific waveguide studied is chosen for its high transmission, an SPDC spectrum with few sidelobes indicating relatively low thickness variations and high brightness at 1550 nm. Once the waveguide is selected, the precise pump wavelengths for the two down-converters are effectively fixed by its quasi-phase matching conditions. Total transmissions of 8.5 dB and 10.2 dB are measured at 796 nm and 408 nm respectively, suggesting high coupling efficiencies and low propagation losses. The downstream setup after the chip is adjusted according to whether the full cascaded performance of the waveguide or an individual down-converter is being studied, and has been described in the following subsections.

Cascaded spontaneous parametric down-conversion (CSPDC) experiment

The true triplet generation efficiency in a cascaded device can be inferred by detecting the daughter photon which is unused in the second SPDC process and either one of the granddaughter photons, and is given by

$$G_{triplet} = \frac{R_{cc}}{\eta_{836}\eta_{1550}P_{pump}} \tag{S1}$$

Here, $R_{cc}$ is the measured coincidence rate between the two photons, $\eta_\lambda$ is the overall detection probability of a photon of central wavelength $\lambda$ (nm) in the experimental setup, and $P_{pump}$ is the on-chip pump power in milliwatts. Separate measurements are performed to determine each factor on the right-hand side of Eq. S1. Additional details about the experiment and data analysis involved in computing the triplet generation efficiency for our CSPDC device (Fig. 3 in the main text) are presented in this subsection.

The 408 nm pump beam is focused into the waveguide under study. An aspheric lens with a broadband anti-reflection coating (Thorlabs C660TME-C, NA = 0.52) is used to collect the waveguide output in the CSPDC experiment. Compared to the reflective objective used for achromatic collection in all other measurements described in later subsections, the aspheric lens provides ≈ 34% higher collection efficiency, improving the coincidence count rates. This improvement comes at the cost of chromatic aberration. When the granddaughter beam is collimated, the daughter beam at 836 nm comes to a focus after the aspheric lens. A plano-convex lens is thus placed downstream to recollimate the 836 nm beam. An angle-tuned 3 nm bandpass filter is used to select the herald wavelength, and the filter transmission spectrum is shown in Fig. S2. The effective bandwidth is defined as the width of a unity-transmission (T = 1) flat-top filter that has the same integrated area as the measured transmission spectrum. With this definition, the effective bandwidth is simply the area under the measured curve and is found to be 4.1 nm. This value is used later to determine the herald-photon brightness.

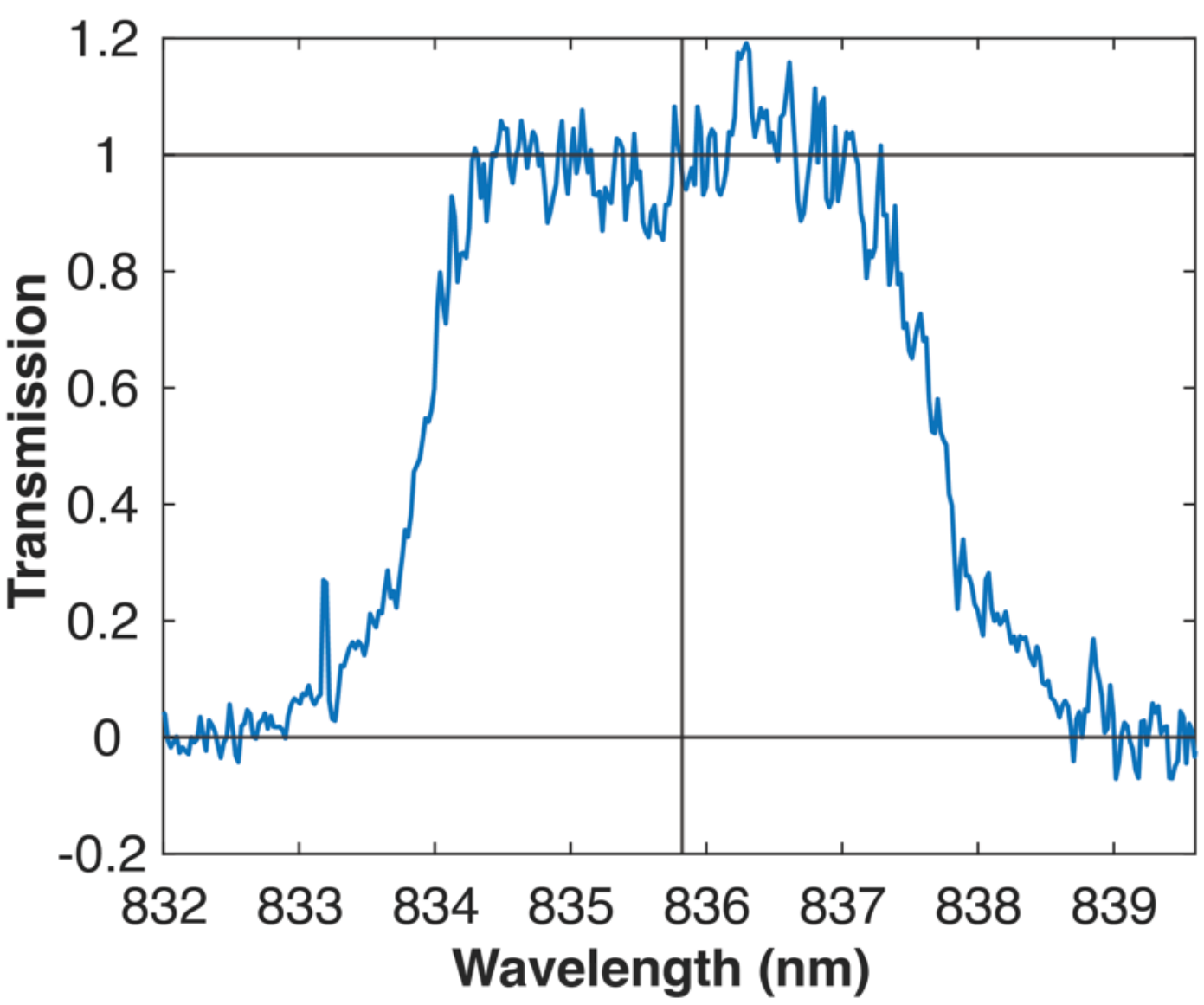

**Fig. S2.** Transmission spectrum of the herald filter. The vertical line denotes the intensity-weighted mean wavelength (first moment) of the transmission spectrum. The filter is centered at 835.87 nm, which is conjugate to 796 nm daughter photons that pump the second down-converter.

Both the daughter and granddaughter beams are coupled into multimode fibers with ≈ 85% coupling efficiency and sent to silicon (NIR detector) and InGaAs (IR detector) single-photon avalanche diodes (SPADs), respectively. Coincidences between the herald 836 nm photon and either of the granddaughter photons are measured as a signature of CSPDC using a time tagger (PicoQuant PicoHarp 300). Coincidence histograms are collected over 1 minute intervals, followed by a measurement of the SPAD detection rates over a 10 second interval. Data is then recorded to a computer over a 19 hour experiment, corresponding to 16 hours of histogram data. A total of 1071811 coincidences are recorded over a 3.6 ns coincidence window. Accidentals from uncorrelated events are integrated over a 60 ns window far from the peak. This value is scaled to a 3.6 ns window to estimate the expected number of accidentals within the coincidence window and subtracted from the raw coincidence data, resulting in a coincidence rate of 0.20 ± 0.03 Hz. The SNR, defined by the ratio of the height of the coincidence peak above the accidental background to the standard deviation of the accidentals, is 7. The pump power is monitored over the entire 19-hour duration by a photodiode that measures 3% reflection of the pump beam from an uncoated wedge, and is shown in Fig. S3A. The average recorded free space pump power is 554 nW. An on-chip pump power of 188 nW is obtained by assuming a 34% input coupling efficiency, which corresponds to half of the total 10.2 dB measured transmission, after accounting for 0.9 dB of propagation loss. The detection probabilities of the daughter and granddaughter photons ($\eta_{836}$ and $\eta_{1550}$) are estimated to be 13.8% and 3.3%, respectively, in separate measurements described in later subsections. Thus, from Eq. S1 the final conversion efficiency of a pump photon to a triplet is determined as 237 ± 36 kHz/mW and the triplet generation rate is 45 ± 7 Hz. A heralding probability of $(6.8 \pm 0.6) \times 10^{-7}$ is obtained by dividing the triplet rate by the measured herald singles rate of $(3 \pm 0.2) \times 10^{5}$ cps. Similar analysis is carried out for different positions of the angle-tuned herald filter to generate the bar chart in Fig. 3D. in the main text.

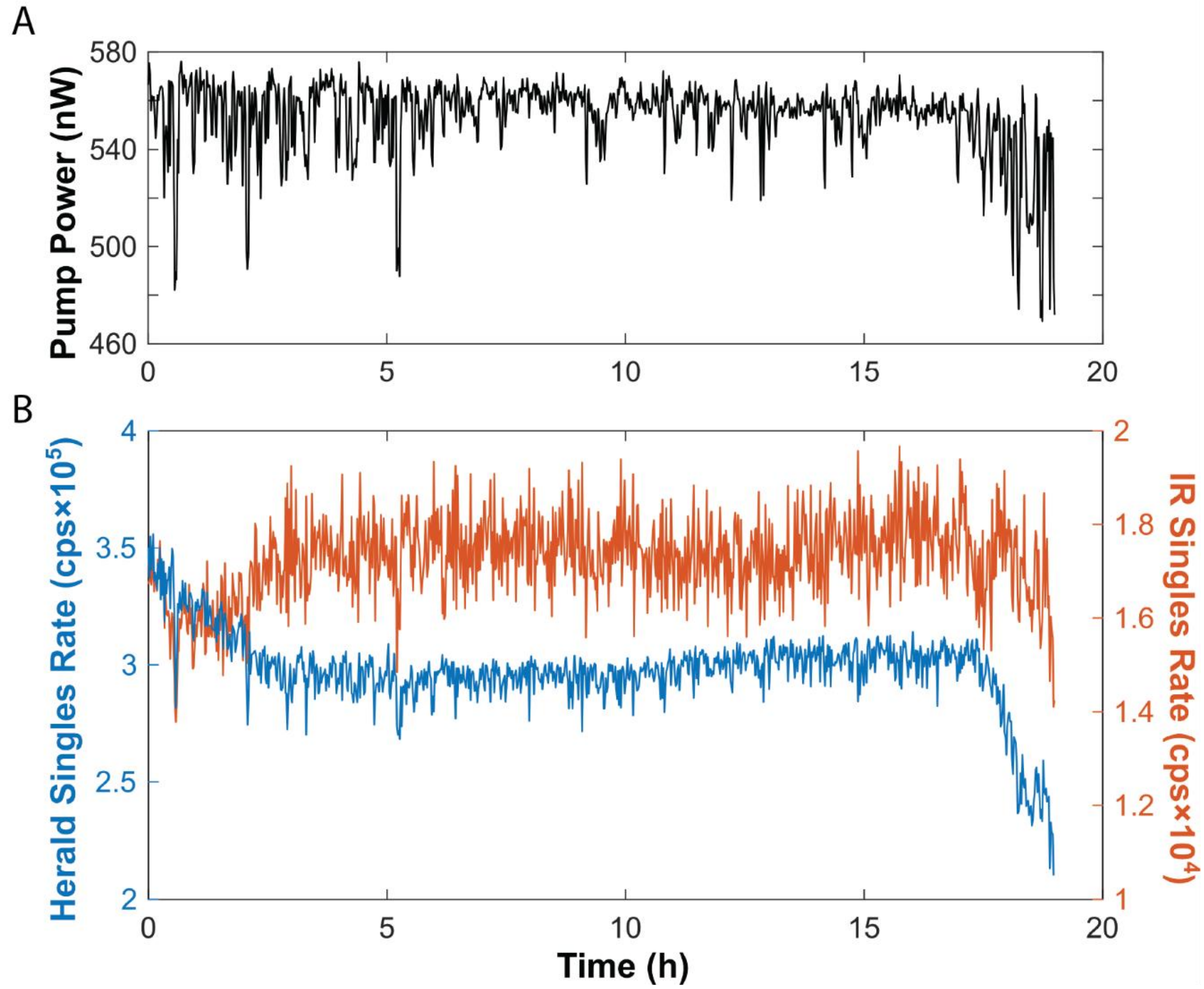

**Fig. S3.** Stability of the CSPDC experiment over time. (A) Pump power and (B) singles counts at the NIR and IR detectors in the experiment. Drifts in the chip coupling of ≈ 15% are seen in the first few hours, as evidenced by the NIR count rate. Drifts in the IR count rate are not as obvious because multiple pump modes are capable of generating IR photons through nondegenerate SPDC.

The CSPDC experiments were challenging due to the high level of background (≈ 12,000:1) of spurious photons relative to granddaughter photons produced through CPSDC. While coincidence detection allows the correlated detection events to be separated from the uncorrelated background events, the probability of detecting an accidental still dominates the probability of detecting a coincidence from CPSDC in our experiments. This necessitated long integration times (> 15 hours) to obtain reasonable signal to noise, even though a CPSDC coincidence is measured every 5 seconds. To determine uncertainties in the coincidence rate due to CPSDC, the uncertainty of the coincidence rate over the detection window needs to be known, in addition to the uncertainty in the accidental rate over this same window. The uncertainties in the accidentals are determined by measuring the mean and standard deviation in the coincidence rate per bin over 0-60 ns, which is a region of the histogram where artifacts due to the histogrammer dead time are not present. The uncertainty in the summed accidental contribution within the coincidence window is then scaled by $\sqrt{N}$, where $N$ is the number of bins in the coincidence window (14 bins of 256 ps duration here). The uncertainty in the total number of coincidences in the window is calculated using standard Poissonian statistics, $\sigma = \sqrt{RT}$, where $\sigma$ is the uncertainty, $R$ is the coincidence rate, and $T$ is the integration time of the experiment. The uncertainty in the coincidence rate of true coincidences is then determined by adding these uncertainties in quadrature.

Running long experiments without a way to correct for chip coupling drifts necessitated a stable experiment. The chip coupling setup is located on an optical table with air bearings to minimize vibrations, and experiments were conducted overnight when foot traffic in the room is minimized. A cardboard box around the setup reduces stray light, air currents, and temperature fluctuations. The stages housing the input and output lenses have been found to be very stable over long periods of time, but the chip position has been found to drift over hour-long time scales. One important source of drift is the cable connected to the crystal oven on which the chip is mounted, which was clamped to the table without introducing too much strain in the connector. Locking the chip stages with set screws and waiting several hours for the setup to stabilize was found to minimize drift to a level acceptable for the experiments done here. The setup stability is confirmed by the recorded singles rate at the two SPADs over the 19-hour acquisition time (Fig. S3B).

Characterization of the first down-converter

In this section, we describe the measurement of the detection probability of the herald photon in the CSPDC experiment and the brightness of the first down-conversion stage. The on-chip spectral brightness of a photon of center wavelength $\lambda$ nm ($B_\lambda$) can be estimated as

$$B_\lambda = \frac{R_\lambda}{\eta_\lambda BW_\lambda P_{pump}} \tag{S2}$$

where $R_\lambda$ is the singles rate of the photon measured on a SPAD and $BW_\lambda$ is the spectral bandwidth, set by the bandwidth of the filter on that measurement arm. The detection probability ($\eta_{836}$) can be determined without knowing the outcoupling efficiency or the transmission of each individual downstream component. Instead, we use the fact that SPDC photons are always generated in pairs obeying energy conservation with respect to the pump. Fig. S4A shows a calibration setup to estimate $\eta_{836}$ and the spectral brightness of the broadband daughter photons. The chip is pumped at 408 nm. The waveguide output is collected with a UV-enhanced, aluminum-coated reflective objective (Thorlabs LMM40X-UVV, NA = 0.5). The objective's central obscuration reduces the collected SPDC photon flux, but it avoids chromatic aberrations over the SPDC wavelength range. SPDC pairs are split on a broadband non-polarizing 50:50 beamsplitter, coupled into multimode fibers, and detected by two near-IR SPADs (Laser Components). A time tagger measures coincidence histograms between the two SPADs. Including the reflectivities of the coupling optics (mirrors and fiber facets), the SPDC is collected with an efficiency of ≈ 85% in each arm. The SPADs used here have a detection efficiency of ≈ 50% at 836 nm. A bandpass filter (10 nm FWHM) is placed in front of one detector and angle tuned so that its center wavelength is 796 nm, the conjugate wavelength of the 836 nm herald used in the CSPDC experiment. Thus, detection of a 796 nm photon in this arm then heralds the presence of its partner photon at 836 nm on the other

SPAD. The conditional probability that the 836 nm photon is detected, given that its partner in the other arm is detected, thus corresponds to the detection (heralding) efficiency of the 836 nm photon in the setup, and is denoted by $\eta_{836,calib}$. Collecting only a narrow spectral region around the center wavelength ensures that the wavelength dependence of the detector does not bias the efficiency estimate. The fits of the singles and coincidence rates as a function of free-space pump power, corrected for the dead time of the SPADs (90 ns) and dark counts, are shown in Fig. S4B. From the ratio of the slopes of the coincidence rate and the singles rate at the detector with the 796 nm bandpass filter, $\eta_{836,calib}$ is evaluated to be 2.6%.

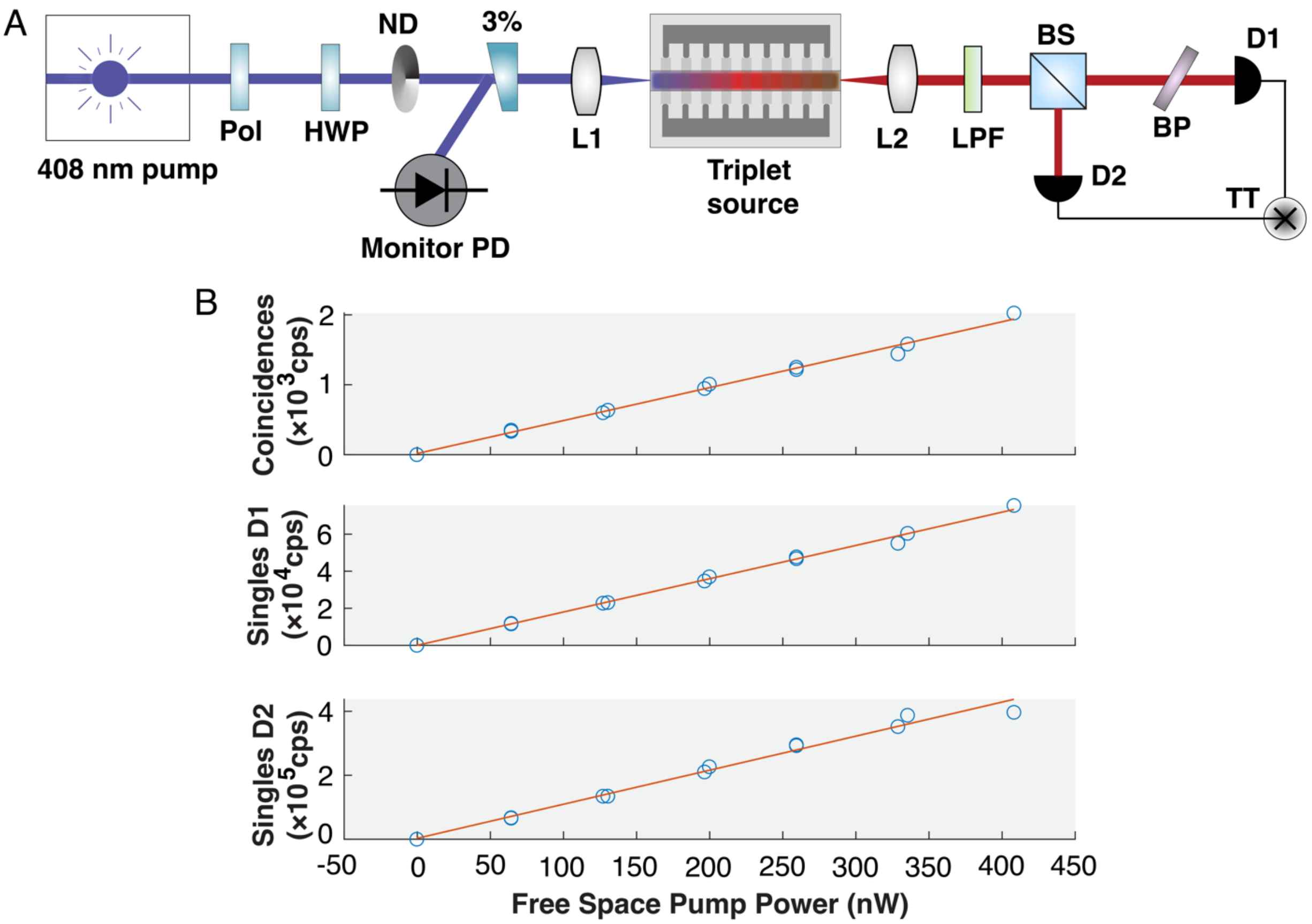

**Fig. S4.** Calibration of $\eta_{836}$. (A) Experimental setup used to estimate the detection probability of 836 nm herald photons. A 796 nm bandpass filter placed before detector D1 heralds the presence of 836 nm photons detected at D2. Abbreviations: Pol, polarizer; HWP, half-wave plate; ND, neutral density wheel; 3%, 3% pickoff wedge; PD, photodiode; L1, aspheric lens; L2, reflective objective; LPF, long-pass filter ($\lambda$ = 500 nm); BS, 50:50 beamsplitter; D1 and D2, Si SPADs; TT, time tagger; BP, 796 nm bandpass filter (FWHM = 10 nm). (B) Coincidence and singles rates at D1 and D2 as a function of free-space pump power (lines are linear fits).

The herald detection efficiency in the CSPDC setup ($\eta_{836,CSPDC}$) is determined by rescaling the calibration result to account for the different collection and coupling optics used in the two experiments. The reflective objective in the calibration setup collects only 34.4% of the light that the asphere used in the CSPDC experiment collects. We characterize the difference due to the rest of the downstream optics in a separate measurement where the Ti:sapphire laser is tuned to 836 nm and directly coupled into the chip in both setups. For each setup, we measure the probability that an 836 nm photon collected at the waveguide output is coupled into the multimode fiber and reaches its detector by taking the ratio of the optical power in the fiber to the power just after the waveguide output coupler lens. The ratio of these probabilities (calibration setup to CSPDC setup) is 0.55. Correcting the measured detection probability of 0.026 by these two factors yields $\eta_{836,CSPDC}$ = 13.8%. The detection efficiency includes propagation loss on the chip, collection efficiency at the lens, output facet reflectivity (17%), fiber coupling efficiency (85%), and SPAD efficiency (50%). By factoring out the known loss contributions in the CSPDC experiment, we deduce a net on-chip and output-coupling loss of about 54% for collection with the aspheric lens. In the CSPDC experiment, we measure $R_{836}/P_{pump}$ = 1.6 GHz/mW from the NIR detector count rate and the estimated on-chip pump power. Using $BW_{836}$ = 4.1 nm, Eq. S2 gives a spectral brightness of 2.8 GHz/mW/nm at 836 nm.

Finally, the SPDC spectrum is recorded with a grating spectrometer and an EMICCD camera (Fig. 2C in the main text). The wavelength axis in the SPDC range (750–880 nm) is calibrated using the tunable 1 MHz linewidth Ti:sapphire laser and a wavemeter, which provides an accurate wavelength reference. The intensity axis is corrected using a reference spectrum of a thermal lamp measured with a calibrated grating spectrometer, and further adjusted for the grating and detector efficiencies. The power spectral density obtained using the EMICCD is normalized to match the experimentally deduced brightness value at 836 nm, resulting in a spectral brightness of 3.5 GHz/mW/nm at 796 nm. The expected spectral brightness and the theoretical spectrum shown in Fig. 2C of the main text are computed from the measured waveguide parameters using the theory in Ref. (*36*). The predicted spectral brightness at 796 nm is 7.7 GHz/mW/nm.

Characterization of the second down-converter

In this section, we characterize (i) the spectral brightness at 1550 nm, (ii) the detection probability of the granddaughter photons, and (iii) the overall pair-generation efficiency of the second SPDC process. Due to the high noise level in the IR detector counts in the CSPDC experiment, the 1550 nm singles rate ($R_{1550}$) cannot be reliably extracted from those measurements. Instead, the second down-converter is pumped independently so that only the photons that originate from SPDC in the second stage are detected (Fig. S5A). The pump beam is coupled into the chip using a high-NA aspheric lens (Thorlabs C230TMD-B, NA = 0.55). The waveguide output is collected using the reflective objective to efficiently collect the ultrabroadband SPDC, which is subsequently split on an ultrafast broadband 50:50 beamsplitter. Both reflected and transmitted beams are collected into multimode fibers, each connected to an InGaAs SPAD. The IR detectors have a detection efficiency of 20% at 1550 nm, measured relative to a calibrated photodiode. Two 1300 nm long-pass filters placed in front of the fiber couplers block the residual pump.

First, $R_{1550}$ is measured when the waveguide is pumped at 796 nm. A 30 nm bandpass filter centered at 1550 nm is placed in front of one of the SPADs (D1). Here, we assume that the SPDC spectral brightness is approximately flat over the 30 nm passband. The rate of 1550 nm singles is recorded as a function of on-chip pump power, which is estimated assuming 45% input coupling, equal to half the measured transmission after correcting for 1.6 dB facet reflections. A linear fit to the singles rate versus pump power yields a slope of $6.4 \times 10^7$ cps/mW. Second, we determine the overall detection probability of an on-chip 1550 nm photon ($\eta_{1550}$) using a calibration scheme similar to that used for the first down-converter. For this calibration, we move the 1550 nm bandpass filter to the opposite arm and use the SPAD on that arm (D2) as the herald. We again assume that the chip transmission and the setup transmission are approximately constant over the 30 nm passband. The coincidence-to-singles ratio then gives the detection probability of the original 1550 nm measurement arm (with detector D1). In this step, the chip is pumped at 775 nm (instead of 796 nm) so that the SPDC is degenerate at 1550 nm. Operating at the degenerate point ensures that a photon transmitted by the 1550 nm bandpass has a partner at 1550 nm in the other arm, and the measured coincidence-to-singles ratio directly yields the heralding efficiency at 1550 nm. Since, the waveguide is not optimally phase matched for degenerate SPDC from 775 nm to 1550 nm, the singles and coincidence counts are significantly low than in the 796 nm pumping case; however, the detection efficiency remains unchanged since it is only a function of the experimental setup at the wavelength of interest. We define the detection efficiency of 1550 nm photons in the calibration setup as $\eta_{1550,calib}$. The singles and coincidence rates as the pump power is varied are shown in Fig. S5B. From the linear fits to this data, we determine $\eta_{1550,calib} = 0.71\%$. Using Eq. S2 with $BW_{1550}$ = 30 nm, a spectral brightness of 0.30 GHz/mW/nm is inferred. Applying the same theoretical analysis as for the first down-converter gives an expected spectral brightness of 0.33 GHz/mW/nm.

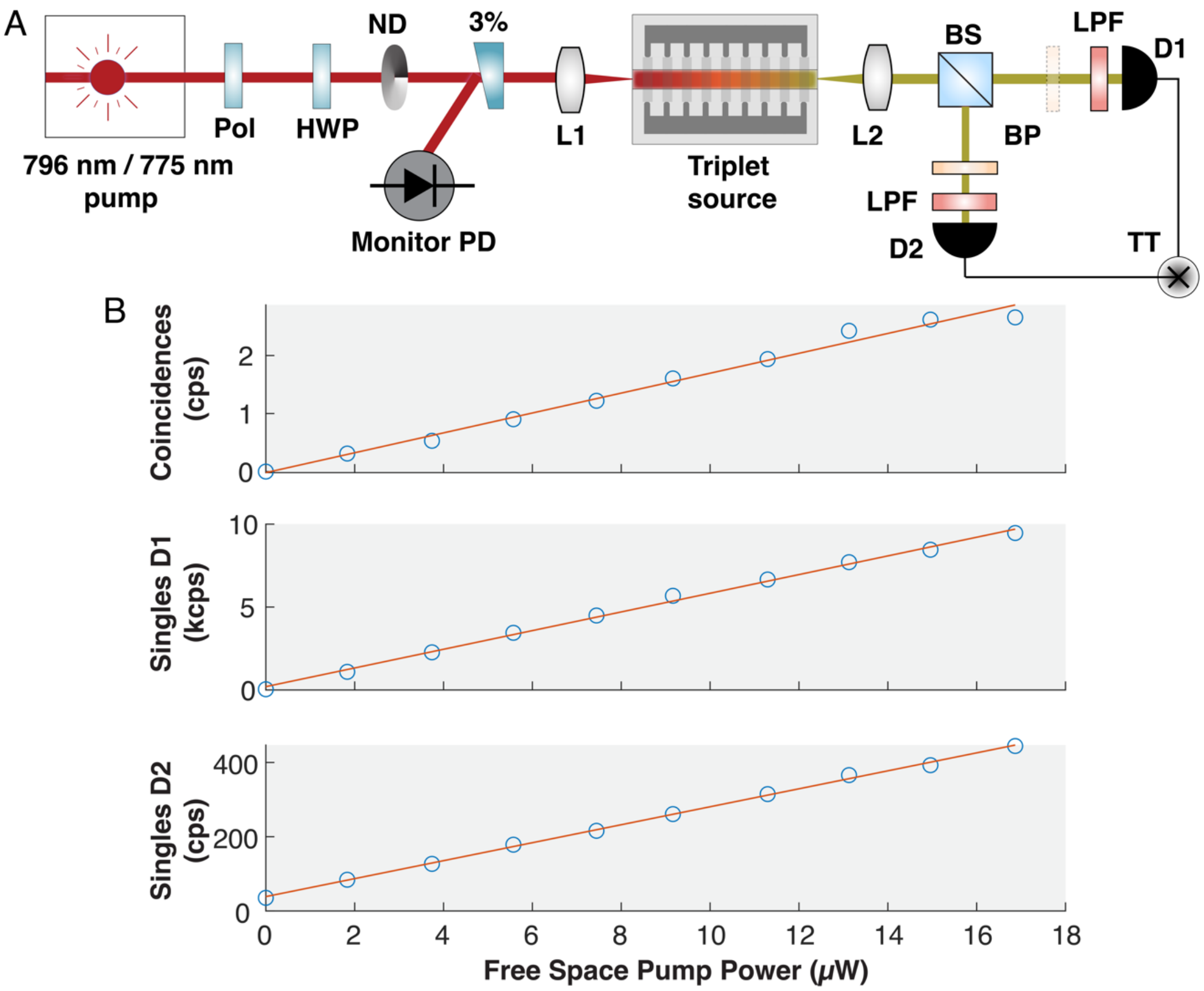

**Fig. S5.** Brightness of the second down-converter and calibration of $\eta_{1550}$. (A) Experimental setup to characterize the performance of the second down-converter. The chip is first pumped at 796 nm, and a 1550 nm bandpass filter (BP) is placed before detector D1, and singles at D1 are recorded. Next, the chip is pumped at 775 nm and the BP is moved to the arm with D2 to determine the detection probability of 1550 nm photons in the arm with D1. Abbreviations: Pol, polarizer; HWP, half-wave plate; ND, neutral density wheel; 3%, 3% pickoff wedge; PD, photodiode; L1, aspheric lens; L2, reflective objective; LPF, long-pass filter; BS, 50:50 beamsplitter; D1 and D2, InGaAs SPADs; TT, time tagger; BP, 1550 nm bandpass filter (FWHM = 30 nm). (B) Coincidence and singles rates at D1 and D2 as a function of free-space pump power when the chip is pumped at 775 nm.

We estimate the detection efficiency in the CSPDC experiment, $\eta_{1550,CSPDC}$ = 3.3% after correcting for the asphere's 43% higher collection efficiency at 1550 nm and the 50:50 beamsplitter used in the calibration setup. Accounting for the known sources of loss in the setup, we estimate a combined on-chip transmission and collection efficiency of 25%. We note that knowledge of the SPAD sensitivity over the full infrared bandwidth is required to accurately estimate the detection probability for broadband granddaughter photons in the CSPDC experiment, where no 1550 nm bandpass filter is used. The SPAD efficiency is expected to decrease at longer wavelengths, so our present estimate likely overestimates the detection probability for these granddaughter photons and therefore underestimates the on-chip triplet generation efficiency in the CSPDC configuration.

The pump acceptance spectrum of the down-converter is measured by changing the pump wavelength and measuring the SPAD counts with the 1550 bandpass (Fig. 3D in the main text). The acceptance bandwidth is estimated at 4.1 nm, which agrees well with the 3.7 nm theoretical value. Finally, the spectrum of the second down-converter is measured. The waveguide is again pumped at 796 nm. The SPDC is collected in a single mode fiber after filtering with a 1300 nm longpass filter and sent to an optical spectrum analyzer (OSA) (Yokogawa AQ6374). A contour plot of the SPDC spectra as a function of pump wavelength is shown in Fig. S6A. Similar to the first down-converter, the measured spectrum is scaled to match the experimentally determined brightness value at 1550 nm. The OSA sensitivity drops at 1700 nm thus cutting off any SPDC longer than that. Energy conservation is used to theoretically reconstruct the entire SPDC spectrum (Fig. S6B) and a 3-dB bandwidth of 116 THz is inferred. The efficiency of the process is determined by integrating the spectrum from 1300 nm to 1592 nm (degenerate wavelength) and doubling it to account for the idlers to be 175 GHz/mW. This corresponds to a down-conversion efficiency of $4.4 \times 10^{-5}$ per pump photon.

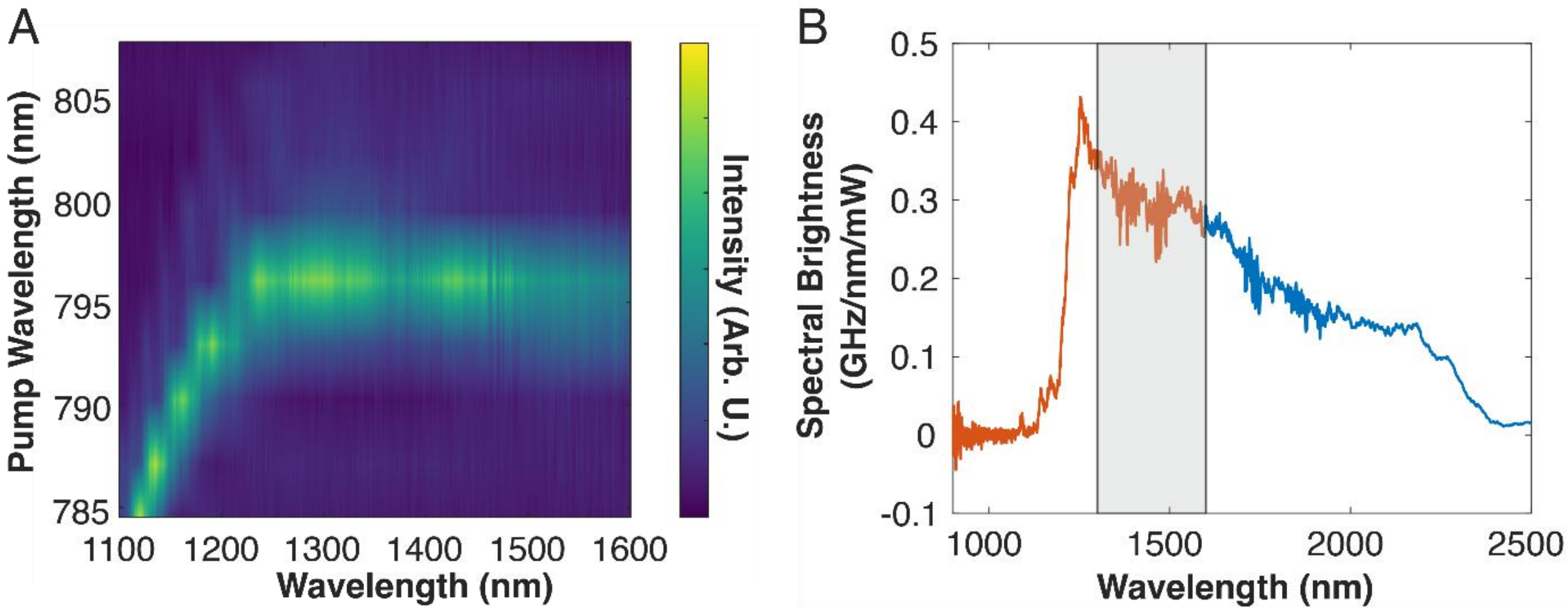

**Fig. S6.** Second down-converter OSA spectral measurements. (A) Measured SPDC spectrum as a function of pump wavelength. (B) Measured signal (orange) and simulated idler (blue) power spectral density. The gray region indicates the detection window used in the CSPDC experiment, extending from a cut-off wavelength of 1300 nm set by the long-pass filters to approximately 1600 nm, limited by the InGaAs detector.

**Origin of accidentals in the CSPDC experiment**

The primary limitation in measuring herald–granddaughter coincidences is the high accidental coincidence rate, caused by noise photons originating in the waveguide. In a monolithic device, any photon that reaches the output and falls within the granddaughter SPAD sensitivity range will be detected, so the measured singles include both the true photons and any noise photons. To investigate the origin of the spurious counts detected by the granddaughter SPAD (1300 nm - 1700 nm), the SPDC in the visible conjugate wavelength range (536 nm - 594 nm) was imaged using a grating spectrometer and an EMICCD detector. Fig. S7 shows the visible SPDC spectrum as the input lens is translated along the plane of the chip. The peaks observed in this band are much weaker as compared to the main daughter-SPDC peak. However, these spurious photons propagate through the waveguide with propagation loss as the main source of attenuation. Thus, their strength is significant when compared with the granddaughter photons that are ~$10^5$ times weaker than the daughter photons. There are three main peaks around 550 nm, 572 nm, and 583 nm, however, additional peaks may appear with longer camera integration times. The corresponding conjugate idler wavelengths are expected near 1580 nm, 1423 nm, and 1359 nm, and can therefore contribute to the noise counts detected by the granddaughter SPAD. The latter two idler wavelengths are expected to be more parasitic since the IR SPAD efficiency is higher at shorter wavelengths. Although the exact pump mode responsible for each peak is hard to determine with a single axis of spatial modal information, the spatially separated lobes indicate that higher order pump modes are getting phase matched to generate these peaks.

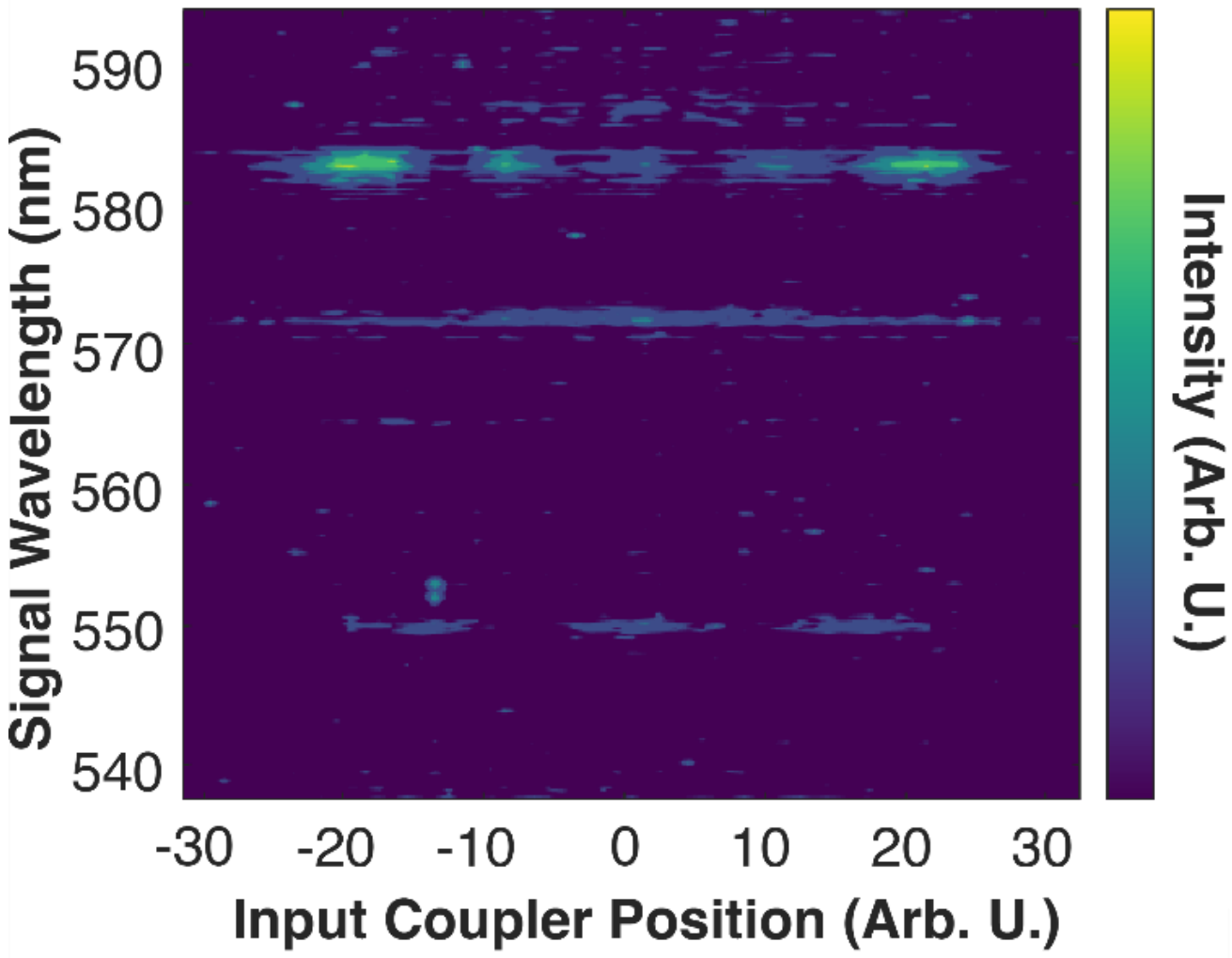

**Fig. S7.** SPDC spectrum of the device in the visible band corresponding to partner photons in the granddaughter SPAD range, when pumped at 407.8 nm. Spectra are measured at different horizontal locations of the input coupler with arbitrary scaling. The input coupler position at 0 denotes the approximate center of the waveguide.

To narrow down the modes involved in the SPDC and the effective poling period producing it, the polarization of the involved pump and SPDC modes is verified to be quasi-transverse electric (quasi-TE) using a polarizer. Next, the supported quasi-TE modes are simulated using the waveguide geometry measured by atomic force microscopy, in an eigenmode solver. All combinations of quasi-TE pump, signal, and idler modes in the wavelength ranges of interest are considered. This analysis shows that higher-order modes can be phase matched by the poling in the first down-conversion region. For example, the fundamental pump mode at 408 nm (TE0), a higher-order signal mode near 572 nm (TE2) and the fundamental idler mode (TE0) satisfy quasi-phase matching for a poling period of ~2.1 μm, matching the first down-converter. Similarly, for the other two peaks, two possible combinations are found that involve higher-order waveguide modes and potentially align with the observed horizontal pump mode profile. However, a unique identification of modes for each peak would require additional information about which specific pump mode is being excited under the different coupling conditions. Overall, the observed phase-matched wavelengths point to the poling period of the first down-converter, indicating that the spurious SPDC is generated before the second region. This could be mitigated in future designs by inserting wavelength-selective filters (e.g., WDMs) between the two sections.

**Pumping with broadband daughter photons**

The daughter SPDC that pumps the second down-converter is centered at 796 nm and has a spectral bandwidth of 4.1 nm, as derived above. The efficiency of generated pairs detected within a given spectral window (1300–2054 nm in the CSPDC experiment) depends on the overlap between the phase-matching response of the second down-conversion process and the pump lineshape. A finite pump bandwidth reduces the overall conversion efficiency compared with monochromatic pumping at the optimal wavelength. Fig. S8 shows the calculated pair-generation efficiency of the second down-converter versus pump wavelength for a single-frequency pump. As shown in the plot, the efficiency drops on either side of degenerate pumping. Thus, for the same total pump power, a reduced efficiency is expected because only a fraction of the pump power overlaps with the optimal phase matching condition, while wavelengths away from this condition contribute less to pair generation. Using the measured spectrum of the 796 nm daughter photons (the conjugate spectrum to Fig. S2) as the pump lineshape, we calculate the expected pair-generation efficiency of down-converter 2 under broadband pumping. Relative to monochromatic pumping at 796 nm with the same total power, the efficiency is 0.83. This factor is used to estimate the expected triplet generation rate in the main text.

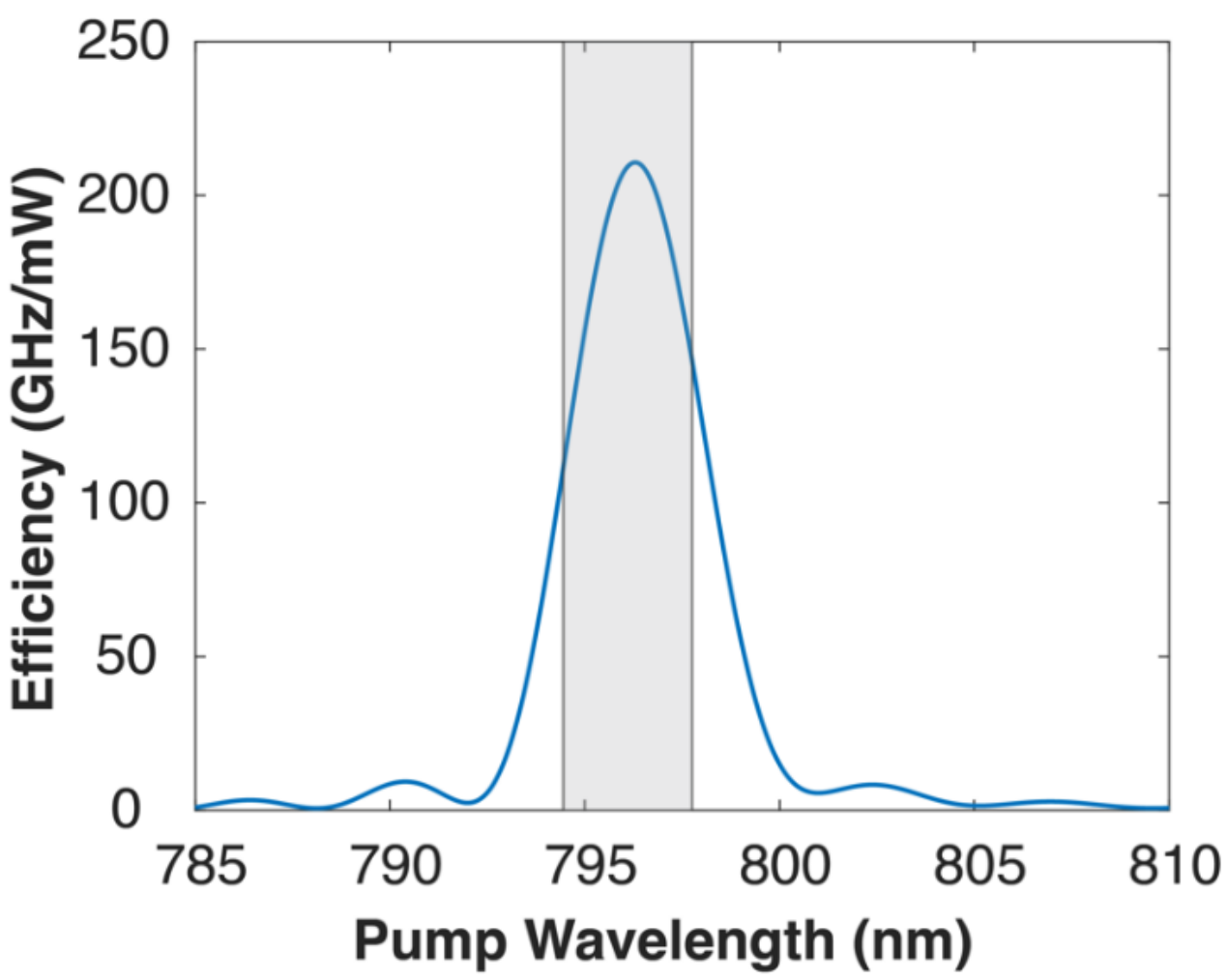

**Fig. S8.** Theoretical single-frequency SPDC efficiency while varying the pump wavelength.